\documentclass[aps,amsmath,amssymb,reprint,twocolumn,pra,superscriptaddress,notitlepage,showpacs,tightenlines]{revtex4-1}
\usepackage{graphicx}
\usepackage{dcolumn}
\usepackage{bm}
\usepackage{siunitx}
\usepackage{color}
\usepackage{amsmath}
\usepackage{float}
\usepackage{verbatim}
\usepackage{soul}
\usepackage[colorlinks,citecolor=blue,linkcolor=red,hyperindex,CJKbookmarks]{hyperref}

\definecolor{pink}{rgb}{0.7922, 0.3373, 0.4118}

\begin{document}

\title{Quantum-Classical Computing for Time-Dependent Ion-Atom Collision Dynamics: Applications to Charge Transfer Cross Section Simulations}

\author{Minchen Qiao}
\affiliation{School of Integrated Circuits, Tsinghua University, Beijing 100084, China}
\affiliation{Frontier Science Center for Quantum Information, Beijing, China}

\author{Yu-xi Liu}
\email{yuxiliu@mail.tsinghua.edu.cn}
\affiliation{School of Integrated Circuits, Tsinghua University, Beijing 100084, China}
\affiliation{Frontier Science Center for Quantum Information, Beijing, China}

\date{\today}

%%%%%%%%%%%%%%%%%%%%%%%%%%%%%%%%%%%%%%%%%%%%%%%%%%%%

\begin{abstract}
The simulation of ion-atom collisions remains a formidable challenge due to the complex interplay between electronic and nuclear degrees of freedom. We present a hybrid quantum-classical computing framework for simulating time-dependent ion-atom collision dynamics, within which two variational quantum time evolution algorithms are implemented. To validate our framework, we simulate the charge transfer dynamics and compute the corresponding cross sections for the proton-hydrogen collision system across an energy range of 1--25~keV. Our results accurately reproduce the charge transfer dynamics with high fidelity and exhibit very good agreement with available experimental and theoretical cross section data across the entire energy range. These results highlight the accuracy and applicability of our hybrid quantum-classical framework for scattering cross section calculations. Our work demonstrates an effective approach for mapping time-dependent many-body collision problems onto near-term quantum computing devices, and also provides promising directions for practical applications of universal quantum computing in the noisy intermediate-scale quantum era.
\end{abstract}
\maketitle

%%%%%%%%%%%%%%%%%%%%%%%%%%%%%%%%%%%%%%%%%%%%%%%%%%%%

\section{Introduction}

Universal quantum computing provides a natural and efficient platform for simulating complex systems and demonstrates provable quantum speedup over classical approaches in specific tasks, such as integer factorization~\cite{shor1994algorithms}, unstructured search~\cite{grover1996fast}, and linear systems solutions~\cite{harrow2009quantum}, all of which require fault-tolerant quantum computers. However, in the current noisy intermediate-scale quantum (NISQ) era, quantum devices remain constrained by limited qubit numbers, gate infidelities, short coherence times, and the absence of fault-tolerant error correction~\cite{preskill2018quantum, bharti2022noisy, lau2022nisq1, cheng2023noisy}. These technical limitations have motivated the rapid development of hybrid quantum-classical algorithms that aim to leverage quantum resources efficiently under these hardware constraints.

Hybrid quantum-classical algorithms~\cite{endo2021hybrid} are computational paradigms that strategically combine quantum and classical resources. In typical workflows, quantum devices handle state preparation, circuit evolution, and measurements, while classical processors manage problem encoding, parameter updates, and data analysis. Representative hybrid approaches like the variational quantum eigensolver (VQE)~\cite{peruzzo2014variational, mcclean2016theory, tilly2022variational} have been successfully applied to chemistry~\cite{peruzzo2014variational, cao2019quantum, mcardle2020quantum, hariharan2024modeling}, materials science~\cite{babbush2018low, bauer2020quantum, oftelie2021simulating, yoshioka2022variational}, biology~\cite{outeiral2021prospects, marchetti2022quantum, baiardi2023quantum}, and drug discovery~\cite{cordier2022biology, bonde2023future, chow2024quantum}. While these achievements primarily address static problems, such as electronic structure calculations~\cite{nam2020ground, tilly2020computation, lotstedt2021calculation, sapova2022variational, innan2024quantum, naeij2024molecular}, combinatorial optimizations~\cite{nannicini2019performance, liu2022layer, amaro2022filtering}, and protein folding~\cite{robert2021resource, uttarkar2024quantum}, many physical systems are governed by time-dependent Hamiltonians, whose dynamics cannot be accurately captured through static approximations. Such dynamic processes are crucial for understanding non-equilibrium quantum physics, including strongly driven quantum materials~\cite{lindner2011floquet, martin2017topological, oka2019floquet, eckhardt2024theory}, ultrafast light-matter interactions~\cite{le2020theoretical, stokes2021ultrastrong, santos2024time}, nuclear reaction pathways~\cite{koonin1977time, guo2007boost, dumitrescu2018cloud, du2021quantum}, as well as ion-atom collisions. These dynamic problems present both challenges and opportunities for quantum simulations.

Ion-atom collision processes~\cite{eichler2005lectures, bransden1992charge, delos1981theory} constitute a fundamental class of time-dependent quantum problems, where electron capture, excitation, and ionization are driven by nuclear motion and described by time-dependent Hamiltonians. These processes provide essential diagnostics for fusion plasmas~\cite{fonck1984determination, isler1994overview, donne2007diagnostics, mcdermott2018evaluation}, enable precise dose delivery in ion-beam cancer therapy~\cite{nikjoo2006track, belkic2014role, durante2010charged}, and play a central role in astrophysical X-ray emissions~\cite{cravens1997comet, kharchenko2000spectra, cravens2002x}. Representative methods for simulating ion-atom collisions include, e.g., atomic orbital close-coupling (AOCC)~\cite{ludde1983method, fritsch1991semiclassical, liu2010radiative, agueny2019electron} and molecular orbital close-coupling (MOCC)~\cite{kimura1989low, zygelman1992charge, harel1998cross}. However, these methods scale exponentially in computational complexity with the number of electrons due to the combinatorial growth of coupled channels. These computational limitations prevent the application of traditional methods to many-electron systems, highlighting the potential of quantum computing to address these exponential scaling bottlenecks.

In this paper, we show an approach to simulate time-dependent many-body collision problems  via a hybrid quantum-classical computing framework. We develop a systematic procedure that maps time-dependent many-body Hamiltonians to a suitable form for digital quantum simulation, and implement two variational quantum time evolution algorithms~\cite{di2024quantum}. Using the proton-hydrogen system as a prototype case, we demonstrate that our hybrid quantum-classical framework can accurately simulate charge transfer dynamics and cross sections via fewer than five qubits. By introducing variational fidelity bound heatmaps, we demonstrate the broad applicability and reliability of our approach across the entire parameter regimes of collision energies and impact parameters. Furthermore, our charge transfer cross sections demonstrate very good agreement with experimental measurements throughout the investigated energy range (1--25~keV). Comparisons with the experimental data reported by McClure~\cite{mcclure1966electron} and Gealy and van Zyl~\cite{gealy1987cross} yield an average relative error of approximately 6\%, with maximum relative errors below 12\%. These results provide compelling evidence for the viability of variational quantum algorithms in addressing practical time-dependent physical problems on NISQ devices.

The paper is organized as follows. In Sec.~\ref{sec:II}, the many-body collision model and its second quantization are briefly summarized. In Sec.~\ref{sec:III},  we first show how to encode time-dependent many-body collision dynamics onto universal quantum computers, and we then tailor two variational quantum time evolution algorithms for our studies on the many-body collision model. In Sec.~\ref{sec:VI}, we demonstrate the practical implementation of quantum simulation on collision dynamics by using an example on proton-hydrogen collisions. The numerical results and performance analysis for charge transfer dynamics and cross sections of the proton-hydrogen collision using both variational algorithms are presented. The comparisons of results to experimental data are given. Finally, in Sec.~\ref{sec:V}, potential improvements to our approaches and their extension to more complex collision systems and other time-dependent many-body problems are discussed, and the conclusions are given. To maintain a focused presentation while ensuring completeness, detailed mathematical derivations are included in the Supplemental Material~\cite{SM}.

%%%%%%%%%%%%%%%%%%%%%%%%%%%%%%%%%%%%%%%%%%%%%%%%%%%%

\section{Basic theoretical model}\label{sec:II}

%%%%%%%%%%%%%%%%%%%%%%%%%%%%%%%
\begin{figure}
    \centering
    \includegraphics[width=1\linewidth]{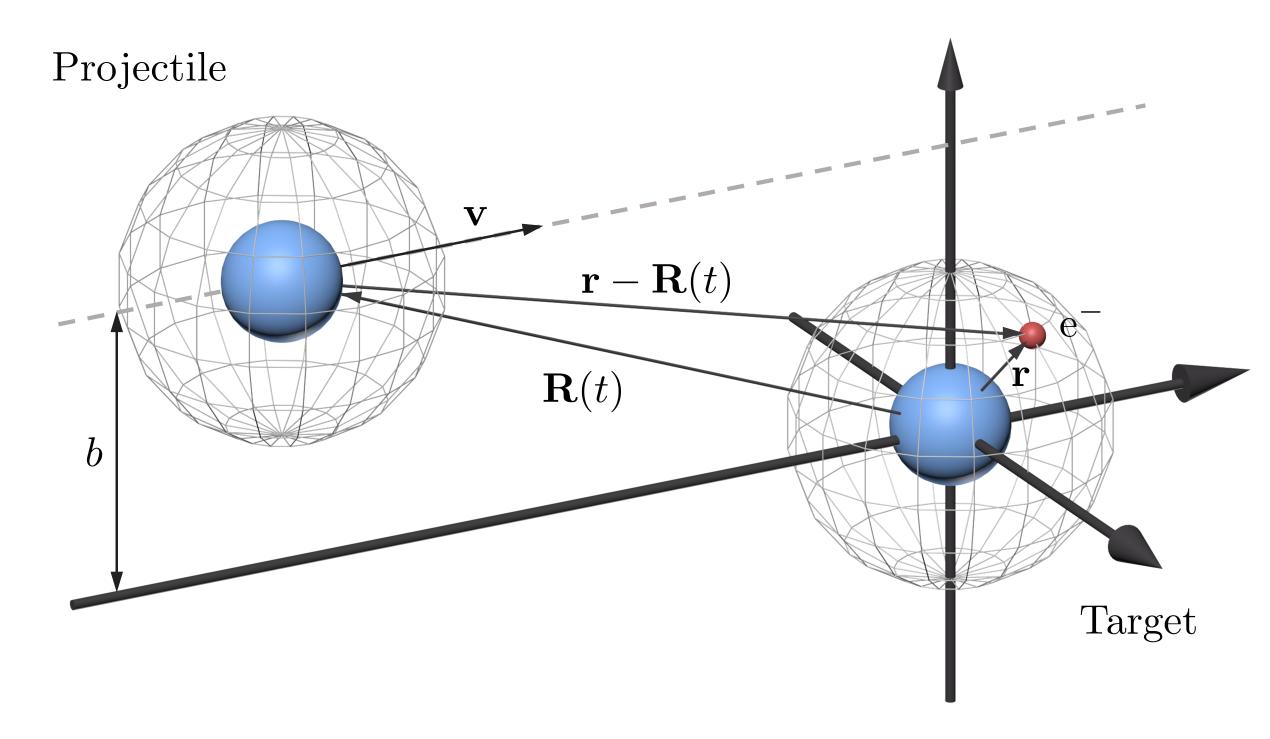}
        \caption{Schematic diagram for a typical ion-atom collision system. The target is fixed at the origin, while the projectile follows a straight-line trajectory $\mathbf{R}(t) = \mathbf{b} + \mathbf{v}t$, with the impact parameter $b$ and the velocity $\mathbf{v}$. The vector $\mathbf{r}$ denotes the position of one of the electrons, e.g.,  in the target. This model illustrates the fundamental time-dependent dynamics that can be extended to more complex collision systems with appropriate modifications for different nuclear and electronic structures.}
    \label{fig: model}
\end{figure}

For completeness of the paper and readers who might not be familiar with ion-atom collisions, in this section, we first briefly summarize the basic theoretical model of time-dependent many-body collision systems. Then, we reformulate this model through second quantization, expressing the system Hamiltonian in terms of fermionic creation and annihilation operators.

\subsection{Time-dependent many-body collision model}

The ion-atom collision model considers a projectile ion (or atom) incident on a stationary target atom (or ion). The relative nuclear motion creates time-dependent Coulomb fields that drive electronic transitions, leading to charge transfer, excitation, and ionization processes. We consider a collision system with $N$ electrons. Within the Born-Oppenheimer approximation~\cite{szabo1996modern}, the time-dependent electronic Hamiltonian in atomic units is expressed as:
\begin{equation}
\begin{aligned}
    H(t) = \sum_{i=1}^N\left(-\frac{1}{2}\nabla_i^2 - \frac{Z_\mathrm{T}}{|\mathbf{r}_{i\mathrm{T}}|} -\frac{Z_\mathrm{P}}{|\mathbf{r}_{i\mathrm{P}}(t)|} \right) + \sum_{i<j}^N \frac{1}{|\mathbf{r}_{ij}|},
\end{aligned}
\label{eq:H2}
\end{equation}
with nuclear charge numbers $Z_\mathrm{T}$ and $Z_\mathrm{P}$ corresponding to the target (T) and the projectile (P), respectively. Here, we define $\mathbf{r}_{i\mathrm{T}} = \mathbf{r}_{i}$, $\mathbf{r}_{i\mathrm{P}}(t) = \mathbf{r}_{i} - \mathbf{R}(t)$, and $\mathbf{r}_{ij}=\mathbf{r}_{i}-\mathbf{r}_{j}$, where the target is placed at the origin, $\mathbf{r}_i$ denotes the position of the $i$-th electron, and $\mathbf{R}(t)$ denotes the internuclear separation between the projectile and the target. The first term represents the one-body Hamiltonian, comprising electron kinetic energy and time-dependent electron-nuclear attractions; the second term represents the two-body electron-electron repulsion. This time-dependent many-body Hamiltonian establishes the foundation for our quantum computational approach.

As schematically shown in Fig.~\ref{fig: model}, we employ the semiclassical impact-parameter approximation~\cite{bransden1979} to describe the projectile trajectory $\mathbf{R}(t)$, where nuclei move classically while electrons evolve quantum mechanically. This approach is well-established for low to intermediate energy collisions where the nuclear motion is much slower than electronic transitions, allowing the electronic wave function to adapt adiabatically to the changing nuclear configuration. The projectile is usually assumed to follow a straight-line trajectory $\mathbf{R}(t)=\mathbf{b}+\mathbf{v}t$, where $b$ is the impact parameter and $\mathbf{v}$ is the constant relative velocity corresponding to the collision energy $E$. We note that the momentum and kinetic energy of the moving projectile are not explicitly included in the Hamiltonian in Eq.~(\ref{eq:H2}). Since the projectile and target nuclei have fundamentally different motions, their dynamic effects should be appropriately incorporated into their respective wave functions to preserve Galilean invariance.

\subsection{Second quantization}

To reformulate the collision problem into a framework suitable for quantum computing, we first adopt second quantization to express the collision Hamiltonian in the Fock space constructed from a spin-orbital basis. That is, the second-quantized form of the Hamiltonian in Eq.~(\ref{eq:H2}) is written as~\cite{sakurai2020modern, helgaker2013molecular}
\begin{equation}
    H_{\mathrm{sec}}(t) = \sum_{pq} h_{pq}(t) a_p^\dagger a_q + \frac{1}{2} \sum_{pqrs} h_{pqrs}(t) a_p^\dagger a_q^\dagger a_r a_s,
\label{eq:secondQ}
\end{equation}
where $a_p^{\dagger}$ and $a_p$ represent the fermionic creation and annihilation operators, respectively, acting on the spin orbital $\chi_p(\mathbf{x},t)$. These operators obey the fermionic anticommutation relations $\{ a_p, a_q^{\dagger} \} = \delta_{pq} $, and $\{ a_p, a_q \} = \{ a_p^{\dagger}, a_q^{\dagger}\} = 0 $, which inherently includes the antisymmetry of fermionic wave functions.

In Eq.~(\ref{eq:secondQ}), the time-dependent one-body coefficients $h_{pq}(t)$ and two-body coefficients $h_{pqrs}(t)$ are numerically integrated using a suitable spin-orbital basis set $\{ \chi_p (\mathbf{x},t) \}$~\cite{helgaker2013molecular, mcweeny1992molecular}. These integrals link the second-quantized Hamiltonian to the first-quantized Hamiltonian in Eq.~(\ref{eq:H2}) as follows:
\begin{align}
    &h_{pq}(t) = \int d\mathbf{x} \, \chi_p^*(\mathbf{x},t) \left(-\frac{\nabla^2}{2} - \frac{Z_\mathrm{T}}{|\mathbf{r}_{i\mathrm{T}}|}-\frac{Z_\mathrm{P}}{|\mathbf{r}_{i\mathrm{P}}(t)|}\right) \chi_q(\mathbf{x},t), \nonumber \\[10pt]
    &h_{pqrs}(t) = \int d\mathbf{x}_1 d\mathbf{x}_2 \, \frac{\chi_p^*(\mathbf{x}_1,t) \chi_q^*(\mathbf{x}_2,t) \chi_r(\mathbf{x}_2,t) \chi_s(\mathbf{x}_1,t)}{|\mathbf{r}_{12}|},
\label{eq:Integrals}
\end{align}
where $\mathbf{x}=(\mathbf{r},m_s)$ denotes the combined spatial and spin coordinates.

In ion-atom collision problems, the spin-orbital functions appearing in Eq.~(\ref{eq:Integrals}) take the form of a product of a spatial orbital and a spin eigenfunction
\begin{equation}
    \chi_p(\mathbf{x},t) = \phi_{\mu}^\tau(\mathbf{r},t) \sigma(m_s),
\label{eq:spinOrbital}
\end{equation}
where $\phi_{\mu}^{\tau}(\mathbf{r},t)$ represents the spatial orbital and $\sigma(m_s)$ is the spin eigenfunction. Specifically, $\tau \in \{\mathrm{T,\; P}\}$ indicates whether the orbital is centered on the target or projectile nucleus. For electrons, the spin eigenfunction $\sigma(m_s)$ corresponds to either spin-up ($\uparrow$) when $m_s = +1/2$ or spin-down ($\downarrow$) when $m_s = -1/2$. Spin-orbital functions constructed from the same spatial orbital but opposite spins are orthogonal.

The choice of spatial orbitals $\phi_{\mu}^{\tau}(\mathbf{r},t)$ depends on the specific collision system and dynamics of interest. For ion-atom collisions, the fundamentally different nuclear motions require careful treatment to preserve Galilean invariance~\cite{bransden1992charge}. To incorporate these effects, we employ the asymptotic form of traveling atomic orbitals, defined as
\begin{equation}
    \phi_{\mu}^\tau(\mathbf{r},t) = \phi_{\mu}(\mathbf{r}_\tau) F^\tau(\mathbf{r},t) e^{-i\epsilon^{\tau}_{\mu}t}.
\label{eq:TAO}
\end{equation}
Here, $\phi_{\mu}(\textbf{r}_{\tau})$ is the corresponding stationary atomic orbital centered on nucleus $\tau$, where $\mathbf{r}_{\tau}$ denotes the electronic position relative to nucleus $\tau$, with $\mathbf{r}_\mathrm{T}=\mathbf{r}$ and $\mathbf{r}_\mathrm{P}=\mathbf{r}-\mathbf{R}(t)$. The phase factor $F^{\tau}(\textbf{r},t)$, which transforms the stationary orbital basis into the traveling orbital basis, is the plane-wave electron translation factor (ETF)~\cite{bates1958electron}, whose specific form depends on the motion of the central nucleus:
\begin{equation}
    \left\{
        \begin{aligned}
            F^\mathrm{T}(\mathbf{r},t) &= 1, \\
            F^\mathrm{P}(\mathbf{r},t) &= \exp\left(i \mathbf{v} \cdot \mathbf{r} - \frac{i}{2} v^2 t\right).
        \end{aligned}
    \right.
\label{eq:ETF}
\end{equation}
These ETFs incorporate the momentum and kinetic energy contributions from orbitals that move with their respective nuclei, providing a more accurate description of the time-dependent electronic structure. The last phase factor $e^{-i\epsilon^{\tau}_{\mu}t}$ accounts for the phase evolution due to the stationary atomic orbital energy $\epsilon^{\tau}_{\mu}$~\cite{schultz2023data}, ensuring correct asymptotic behavior of the electronic wave function.

After constructing the second-quantized Hamiltonian, the many-body wave functions can be represented using the occupation number representation. For practical simulations, a finite truncation of the active orbital space is necessary to balance accuracy and computational feasibility~\cite{sato2018communication}. When the collision dynamics are confined to $M$ spin orbitals, the many-body electronic state can be expressed in the occupation number representation as
\begin{equation}
    |\psi\rangle_{\mathrm{sec}} = |f_{M-1}, f_{M-2}, \cdots, f_p, \cdots, f_0\rangle ,
\label{eq:fermionState}
\end{equation}
where $f_p=1$ if the spin orbital $\chi_p(\mathbf{x},t)$ is occupied by an electron, and $f_p=0$ if it is empty. Although the occupation number representation provides a compact description compared to explicit Slater determinants, classical computers still require vectors of dimension $2^M$ to store and manipulate these states, which becomes intractable for large systems. By contrast, quantum computers naturally represent these states using only $M$ qubits, offering a scalable computational advantage for many-electron systems.

%%%%%%%%%%%%%%%%%%%%%%%%%%%%%%%%%%%%%%%%%%%%%%%%%%%%

\section{Qubit encoding and quantum algorithms}\label{sec:III}

As a necessary complement for following simulations via universal quantum computing and for readers unfamiliar with quantum computing, in this section, we first show how to encode the time-dependent second-quantized Hamiltonian into a universal quantum computing form via the Bravyi-Kitaev transformation~\cite{bravyi2002fermionic}. Then, we briefly review two quantum algorithms, which are tailored for solving the time-dependent Hamiltonian of ion-atom collisions.

\subsection{Qubit encoding for time-dependent Hamiltonians}

To compute the collision Hamiltonian in Eq.~(\ref{eq:secondQ}) by universal quantum computers, we next adopt the Bravyi-Kitaev (BK) transformation~\cite{bravyi2002fermionic} to map the fermionic creation and annihilation operators onto Pauli operators. The BK transformation has been found to require relatively fewer quantum gates compared to the Jordan-Wigner transformation~\cite{jordan1928paulische}, especially with increasing system size~\cite{tranter2018comparison,  uvarov2020variational}. In the BK transformation, the creation and annihilation operators are defined in terms of Pauli operators as
\begin{equation}
    \begin{aligned}
        a_p^\dagger &= \frac{1}{2} X_{U(p)} \otimes \left(X_p \otimes Z_{P(p)} - i Y_p \otimes Z_{R(p)}\right), \\
        a_p         &= \frac{1}{2} X_{U(p)} \otimes \left(X_p \otimes Z_{P(p)} + i Y_p \otimes Z_{R(p)}\right).
    \end{aligned}
\label{eq:BK}
\end{equation}
Here, the update set $U(p)$, parity set $P(p)$, and remainder set $R(p)$ specify the qubit indices for the required bit-flip and phase operations in the BK transformation. $X$, $Y$ and $Z$ are Pauli operators. These sets vary with the index $p$ and are well defined in previous works, e.g., Refs.~\cite{seeley2012bk,tranter2015bk}, and are also detailed in Sec.~II of the Supplemental Material.

Using the BK encoding, the fermionic Hamiltonian in Eq.~(\ref{eq:secondQ}) can be mapped to its qubit representation, which takes the form of a linear combination of Pauli strings
\begin{equation}
    \begin{aligned}
        {H}_\mathrm{BK}(t) & = \sum_{\gamma}g_{\gamma}(t)H_{\gamma} \\
        & = \sum_{\gamma}g_{\gamma}(t)\bigotimes_{p=0}^{N_Q-1}\sigma_{\gamma,p}.
    \end{aligned}
\label{eq:LCU}
\end{equation}
Here, $\gamma$ indexes the Pauli strings, and the time-dependent expansion coefficients $g_{\gamma}(t)$ are computed from the second quantization coefficients in Eq.~(\ref{eq:Integrals}) and the transformation coefficients in Eq.~(\ref{eq:BK}). The time dependence of these expansion coefficients arises from the evolving internuclear separation $\mathbf{R}(t)$ and the time-varying spin orbitals $\chi_p(\mathbf{x},t)$. The time-independent Hamiltonian terms $H_\gamma$ belong to the set $\mathbb{U}_{N_Q}$ of the $N_Q$-qubit Pauli strings, where each element takes the form:
\begin{equation}\label{eq:U}
   U = \bigotimes_{p=0}^{N_Q-1} \sigma_p ,
\end{equation}
where $p$ labels the qubit index and each $\sigma_p \in \{I, X, Y, Z\}$ acts on qubit $p$. In the context of Eq.~(\ref{eq:LCU}), $\sigma_{\gamma,p}$ denotes the specific Pauli operator acting on qubit $p$ for the $\gamma$-th Pauli string. This Hamiltonian form, known as a linear combination of unitaries (LCU), requires $N_Q = M$ qubits corresponding to the $M$ spin orbitals.

We then use the transformation matrix $\beta_\text{BK}$ to map the occupation number states in Eq.~(\ref{eq:fermionState}) into the corresponding BK qubit representation:
\begin{equation}
    |\psi\rangle_{\mathrm{BK}} = |q_{M-1}, q_{M-2},\cdots, q_p ,\cdots, q_0\rangle, \
\label{eq:qubitState}
\end{equation}
with
\begin{equation}
q_p = \sum_q [\beta_{\mathrm{BK}}]_{pq} f_q.
\label{eq:qubitStateM}
\end{equation}
The definition of the transform matrix $\beta_\mathrm{BK}$ is discussed in Refs.~\cite{seeley2012bk,tranter2015bk} and given in Sec.~II of the Supplemental Material. This transformation maps fermionic states onto qubit states that can be directly prepared and manipulated on quantum hardware. With this qubit encoding, quantum algorithms can efficiently solve the time evolution of collision dynamics that would be challenging for classical methods.

%%%%%%%%%%%%%%%%%%%%%%%%%%%%%%%%%%%%%%%%%%%%%%%%%%%%

\subsection{Variational quantum time evolution algorithms}

For comparisons, we apply two variational quantum time evolution algorithms summarized below for simulating the dynamics governed by the Hamiltonian in Eq.~(\ref{eq:LCU}). Although both algorithms share the same variational framework, they differ in the choice of ansatz and variational parameters, leading to distinct circuit constructions, computational characteristics, and quantum resource requirements. These complementary approaches represent different strategies for balancing simulation accuracy with quantum resource constraints in quantum dynamics simulations, particularly in collision processes. By studying these approaches, we demonstrate how variational quantum time evolution algorithms can be applied to simulate the charge transfer dynamics and cross sections in collision problems using universal quantum computers.

%%%%%%%%%%%%%%%%%%%%%%%%%%%%%%%%%%%%%%%%%%%%%%%%%%%%%
\begin{figure*}
    \centering
    \includegraphics[width=0.85\linewidth]{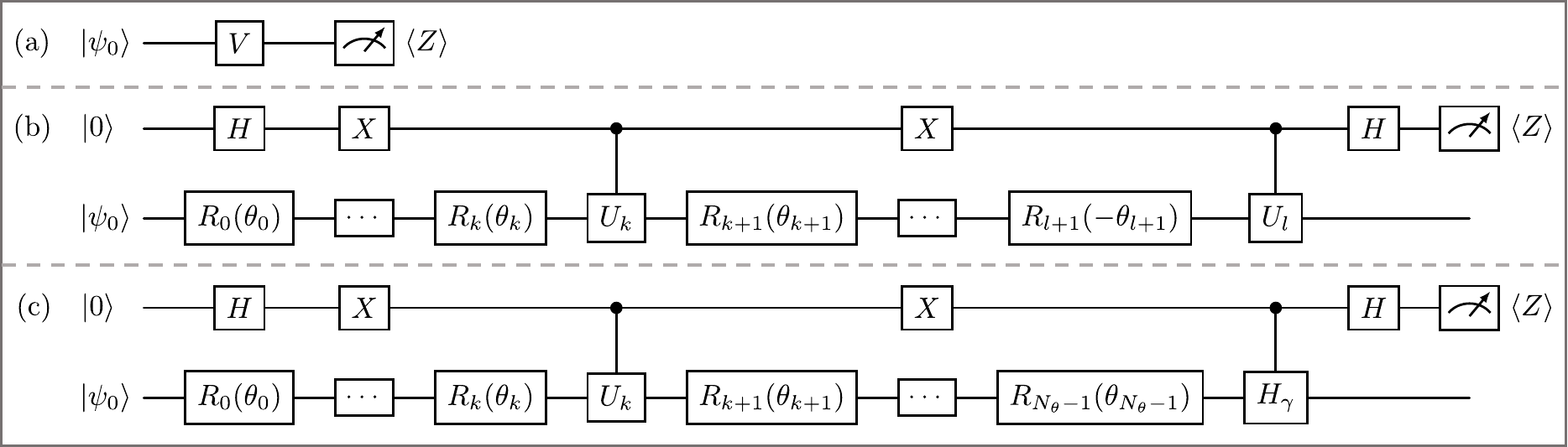}
    \caption{Quantum circuits used in variational quantum time evolution algorithms: (a) direct measurement circuit in the quantum-assisted method. This circuit enables measurement of the time-independent matrix elements $A_{ij}$ and $D_{\gamma,ij}$ required in the QAS parameter evolution equations. (b) and (c) Hadamard test circuits in the quantum-parameterized method, which require one additional ancillary qubit compared to (a). These circuits enable measurement of the time-dependent matrix elements $A^\mathrm{R}_{kl}(t)$ via circuit (b) and $C^\mathrm{I}_{\gamma,k}(t)$ via circuit (c) for the AVQDS parameter evolution equations. In practical implementations, adjacent Hermitian conjugate gate pairs can be removed. We note that $H$ in the circuit denotes a Hadamard gate, not a Hamiltonian.}
    \label{fig: circuits}
\end{figure*}
%%%%%%%%%%%%%%%%%%%%%%%%%%%%%%%%%%%%%%%%%%%%%%%%%%%%%

\subsubsection{Simplified Quantum-Assisted Method}

The quantum-assisted simulator (QAS) algorithm~\cite{bharti2021quantum} is a representative example of the quantum-assisted method~\cite{lau2021quantum,haug2022generalized,bharti2021quantum,bharti2021iterative,lau2022nisq2}. In QAS, the variational ansatz is expressed as a linear combination of quantum basis states from a specifically constructed quantum basis set $\mathbb{CS}^{\prime}_K$, where the combination coefficients are time-dependent and complex:
\begin{equation}
    |\Psi(\boldsymbol{\alpha}(t))\rangle = \sum_{|\varphi_i\rangle \in \mathbb{CS}'_K} \alpha_i(t) |\varphi_i\rangle.
\label{eq:QASansatz}
\end{equation}
Here, the complex parameters $\{\alpha_i(t)\}$ are independent of quantum circuit construction and are processed purely on classical computers, while the basis set $\mathbb{CS}^{\prime}_K$ consists of predefined quantum states that can be readily prepared using quantum circuits.

To construct the quantum basis set $\mathbb{CS}^{\prime}_K$, we first build a Hamiltonian-inspired set $\mathbb{CS}_K$ by merging the LCU Hamiltonian with the Krylov subspace method~\cite{haug2022generalized}. Starting from the $N_{\gamma}$-term LCU Hamiltonian in Eq.~(\ref{eq:LCU}) and an initial state $|\varphi_0\rangle$, we construct the cumulative $K$-moment state set as:
\begin{equation}
    \mathbb{CS}_K = \mathbb{S}_0 \cup \mathbb{S}_1 \cup \cdots \cup \mathbb{S}_K,
\end{equation}
with
\begin{equation}
    \mathbb{S}_K = \{U_{i_K}\cdots U_{i_1}|\varphi_0\rangle\}^{N_\gamma}_{i_1,\ldots,i_K=1},
\end{equation}
where $U_{i_K}\in\{H_\gamma\}^{N_\gamma}_{\gamma=1}$ are Pauli strings given in Eq.~(\ref{eq:LCU}). The multiplication closure property of Pauli strings naturally limits the number of cumulative $K$-moment states. By increasing $K$, we can expand $\mathbb{CS}_K$ to its maximum size. These cumulative $K$-moment states then span a subspace of the Hilbert space that contains all dynamically accessible states under the given Hamiltonian evolution~\cite{bharti2021quantum,bharti2021iterative,lau2022nisq2}.

However, there are many linear dependencies among these cumulative $K$-moment states, as some states differ only by a global phase. To resolve this, we introduce the following equivalence relation~\cite{gunther2007projective}:
\begin{equation}
    |\psi\rangle \sim |\varphi\rangle \text{ if and only if } |\varphi\rangle = e^{i\lambda} |\psi\rangle \text{ for } \lambda \in \mathbb{R},
\label{eq:equivRelation}
\end{equation}
and define a projective space $\mathbb{P}(\mathbb{CS}_K)=\mathbb{CS}_K/\sim$. Thus, we can build the simplified cumulative K-moment set $\mathbb{CS}^{\prime}_K$ by selecting one representative quantum state from each equivalence class in the projective space $\mathbb{P}(\mathbb{CS}_K)$, thereby forming a complete and linearly independent standard orthogonal qubit basis. This construction of the quantum basis set not only ensures the orthogonality of these quantum basis states but also significantly reduces the dimensionality of the basis set while preserving the ability to capture the quantum dynamics of the system accurately.

By applying McLachlan's variational principle~\cite{mclachlan1964variational} to the ansatz of Eq.~(\ref{eq:QASansatz}) and the time-dependent Hamiltonian of Eq.~(\ref{eq:LCU}) (see Sec.~I of the Supplemental Material in detail), we obtain ordinary differential equations (ODEs) for the complex variational parameters as:
\begin{equation}
    \mathbf{A} \dot{\boldsymbol{\alpha}}(t) = -i \sum_\gamma g_\gamma(t) \mathbf{D}_{\gamma} \boldsymbol{\alpha}(t),
\label{eq:EOMqas}
\end{equation}
with the matrix elements $A_{ij}$ and $D_{\gamma,ij}$ defined as:
\begin{equation}
    \begin{aligned}
        A_{ij} &= \langle \varphi_i | \varphi_j \rangle, \\
        D_{\gamma,ij} &= \langle \varphi_i | H_{\gamma} | \varphi_j \rangle,
    \end{aligned}
\label{eq:AijDij}
\end{equation}
where $\mathbf{A}\equiv [A_{ij}] $ is the overlap matrix and $\mathbf{D}_\gamma\equiv [ D_{\gamma,ij}]$ are the coupling matrices for the $\gamma$-th Pauli string. The total time-dependent coupling matrix is then given by $\mathbf{D}(t) = \sum_\gamma g_\gamma(t) \mathbf{D}_{\gamma}$. Quantum computation is required only for evaluating these time-independent matrix elements  $A_{ij}$ and $D_{\gamma,ij}$. Once the quantum measurements have been accomplished, the parameter evolution proceeds entirely on classical computers via
\begin{equation}
    \boldsymbol{\alpha}(t + \delta t) = \boldsymbol{\alpha}(t) + \dot{\boldsymbol{\alpha}}(t) \delta t,
\label{eq:alphaUpdate}
\end{equation}
without requiring quantum-classical feedback. This hybrid paradigm allows flexible time step adjustment for desired accuracy without incurring additional quantum overhead. Furthermore, the basis set simplification enables the resulting ODEs to evolve within a compact and tractable parameter space rather than the exponentially large full Hilbert space.

To implement QAS on quantum computers, we now specify how to measure the required matrix elements using quantum circuits. We assume that measurements are performed in the computational basis. The matrix elements $A_{ij}$ and $D_{\gamma,ij}$ can thus be reduced to expectation value measurements of Pauli strings~\cite{mitarai2019methodology}:
% , corresponding to the global observable $Z=\bigotimes_{p=0}^{N_Q-1} Z_p$
\begin{equation}
    \begin{aligned}
        \langle \varphi_0 | U | \varphi_0 \rangle &= \left\langle \varphi_0\left | \bigotimes_{p=0}^{N_Q-1} \sigma_p \right| \varphi_0 \right\rangle \\
        &=  \left\langle \varphi_0 \left| \bigotimes_{p=0}^{N_Q-1} [V_p^\dagger Z_p V_p] \right| \varphi_0 \right\rangle \\
        &= \langle \varphi_0 | V^\dagger Z V | \varphi_0 \rangle,
    \end{aligned}
\label{eq:directMeasure}
\end{equation}
where $U$ represents a general Pauli string as defined in Eq.~(\ref{eq:U}), derived from the matrix elements $A_{ij}$ and $D_{\gamma,ij}$. Here, $V_p$ denotes the single-qubit rotation that maps $\sigma_p$ to $Z_p$, with $V=\bigotimes_{p=0}^{N_Q-1} V_p$ and $Z=\bigotimes_{p=0}^{N_Q-1} Z_p$ representing the global rotation and measurement operators. This enables direct measurement on the computational basis, as shown in Fig.~\ref{fig: circuits}(a), without requiring Hadamard tests or ancillary qubits.

\subsubsection{Quantum-Parameterized Method}

The quantum-parameterized method, also known as the variational quantum simulation method~\cite{yao2021adaptive, lee2022simulating, miessen2023quantum}, provides a real-parameter implementation of the variational principle. As a representative of this method, we employ the adaptive variational quantum dynamics simulation (AVQDS) algorithm~\cite{yao2021adaptive}, whose variational ansatz is dynamically constructed using parameterized quantum circuits with time-dependent real parameters
\begin{equation}
    |\Psi(\boldsymbol{\theta}(t))\rangle = \prod_k R_k(\theta_k(t)) |\varphi_0\rangle = \prod_k e^{-i \theta_k(t) U_k} |\varphi_0\rangle,
\label{eq:AVQDSansatz}
\end{equation}
where $\{\theta_k(t)\}$ are real parameters that can be directly encoded in quantum circuits through rotation gates ${R}_{k}(\theta_{k})= e^{-i\theta_{k}{U}_k}$, with $U_k$ being a Pauli string selected from a given Pauli operator set. The number of parameters $N_{\theta}$, corresponding to the number of parameterized gates in the ansatz circuit, increases adaptively during the simulation.

This adaptive construction strategy is a feature of AVQDS, where operators from the predefined Pauli string set are iteratively chosen and added to the ansatz to minimize the McLachlan distance ~\cite{yuan2019theory}:
\begin{equation}
    L^2(t) = \left\| \left(\frac{d}{dt} + iH(t)\right) |\Psi(t)\rangle \right\|^2,
\label{eq:L2}
\end{equation}
which represents the squared distance between the parameterized evolving state and the theoretically exact state, and can serve as a measurable and traceable quantity to evaluate the variational evolution accuracy.

At each time step, AVQDS evaluates whether the current ansatz satisfies the threshold criterion $L^2 < L^2_\text{cut}$. If the criterion is not met, each operator in the set is tested by incorporating it into the antasz circuit in the form $e^{-i\theta_{k}U_{k}} |\Psi(\boldsymbol{\theta}(t))\rangle$, with $\theta_k$ initialized to zero. Although this change leaves the quantum state unchanged, it modifies $L^2$ through potentially non-zero derivative terms. The operator that achieves the maximum reduction in $L^2$ is then selected and added to the circuit, increasing the parameter number to $N_{\theta}+1$. The selection process continues iteratively until reaching the threshold criterion $L^2 < L^2_\text{cut}$. If the ansatz requires further expressibility, additional circuit layers can be incorporated, as suggested in Ref.~\cite{zhang2024adaptive}. Such an adaptive mechanism allows for a non-redundant increase in $N_{\theta}$ while maintaining the error of each simulation step within the McLachlan threshold $L^2_\text{cut}$.

Practically, the Pauli operator string set can be constructed in various ways, including using a complete set $\mathbb{U}_{N_Q}$ of $N_Q$-qubit Pauli strings or only considering the Pauli strings generated from the system Hamiltonian. The set size can be reduced based on system symmetries, such as particle number and spin conservation~\cite{yao2021adaptive}. In this work, since the McLachlan distance is evaluated classically and the adaptive nature of AVQDS, we employ the complete set $\mathbb{U}_{N_Q}$ to achieve optimal performance. In quantum hardware implementations where evaluating $L^2$ requires actual measurements, a smaller operator pool composed of one- and two-qubit local gates may be preferred to reduce measurement overhead.

Having established the real-parameterized ansatz, we apply McLachlan's variational principle to obtain the time evolution equations for the variational parameters in the form of real ODEs~(see Sec.~I of the Supplemental Material in detail):
\begin{equation}
    \mathbf{A}^{\mathrm{R}}(t)\, \dot{\boldsymbol{\theta}}(t) = \sum_\gamma g_\gamma(t)\, \mathbf{C}_\gamma^{\mathrm{I}}(t).
\label{eq:EOMavqds}
\end{equation}
Here, the time-dependent matrix elements ${A}^\mathrm{R}_{kl}(t)$ and vector components ${C}^\mathrm{I}_{\gamma,k}(t)$ are defined as
\begin{equation}
    \begin{aligned}
        A^\mathrm{R}_{kl}(t) &= \mathrm{Re}\left(\frac{\partial \langle \Psi(\boldsymbol{\theta}(t)) |}{\partial \theta_k} \frac{\partial | \Psi(\boldsymbol{\theta}(t)) \rangle}{\partial \theta_l}\right), \\
        C^\mathrm{I}_{\gamma,k}(t) &= \mathrm{Im}\left(\frac{\partial \langle \Psi(\boldsymbol{\theta}(t)) |}{\partial \theta_k} H_\gamma |\Psi(\boldsymbol{\theta}(t))\rangle \right).
    \end{aligned}
\label{eq:AklCk}
\end{equation}
All these time-dependent elements and the terms required for evaluating $L^2$ need to be measured on quantum computers, and then transferred to classical computers for threshold check and parameter updates at each iteration step. The parameters are updated according to
\begin{equation}
    \boldsymbol{\theta}(t + \delta t) = \boldsymbol{\theta}(t) + \dot{\boldsymbol{\theta}}(t) \delta t,
\label{eq:thetaUpdate}
\end{equation}
forming a quantum-classical feedback loop.

The matrices $A^\mathrm{R}_{kl}(t)$ and $ C^\mathrm{I}_{\gamma,k}(t)$ can be obtained via $Z$-basis measurement of the ancillary qubit in the Hadamard test~\cite{li2017efficient}, as illustrated in Figs.~\ref{fig: circuits}(b) and (c). Additionally, the adaptive operator selection process in AVQDS requires evaluation of the McLachlan distance $L^2$, which involves expectation values that can be measured using variants of the direct measurement approach of Fig.~\ref{fig: circuits}(a) with additional parameterized gates, as illustrated in Ref.~\cite{yao2021adaptive}. Detailed quantum circuit implementations for both variational quantum time evolution algorithms are provided in Secs.~I and III  of the Supplemental Material.

%%%%%%%%%%%%%%%%%%%%%%%%%%%%%%%%%%%%%%%%%%%%%%%%%%%

\section{Quantum computing for $\mathrm{H}^+ + \mathrm{H}(1s)$ Collisions}\label{sec:VI}

Based on theoretical frameworks presented in Secs.~\ref{sec:II} and \ref{sec:III}, we now use the proton-hydrogen collision system as a benchmark example to validate our hybrid quantum-classical approach. While conceptually simpler than realistic many-electron systems, this model preserves the essential dynamical features of time-dependent charge transfer and nonadiabatic electronic motion that are ubiquitous in ion-atom collision processes. The proton-hydrogen system plays a foundational role in the development of collision theory~\cite{abrines1966classical, kolakowska1998excitation, agueny2019electron} and has demonstrated clear extensibility to multielectron contexts~\cite{gao2019double, jorge2018ionization, murakami2012single}. Here, the system serves as a physically meaningful and computationally tractable platform for benchmarking algorithmic performance, while providing a concrete step toward the simulation of more complex collision processes involving correlated many-electron dynamics.

\subsection{Qubit representation and algorithm implementation for the $\mathrm{H}^+ + \mathrm{H}(1s)$ Collision Model}

The proton-hydrogen collision system exhibits rich physics across different energy regimes. To demonstrate our framework while maintaining computational feasibility, we focus on the low to intermediate energy range (1--25~keV), where the collision dynamics are dominated by the symmetrical resonance charge transfer process~\cite{delos1981theory, bransden1992charge, eichler2005lectures}:
\begin{equation}
    \mathrm{H}^+ + \mathrm{H}(1s) \rightarrow \mathrm{H}(1s) + \mathrm{H}^+,
\label{eq:Reaction}
\end{equation}
which describes the electron motion in the field of two protons. For this collision model, the projectile ion and target atom in Fig.~\ref{fig: model} are replaced by $\mathrm{H}^+$ and $ \mathrm{H}(1s)$, respectively.   The Hamiltonian in Eq.~(\ref{eq:H2}) is reduced to
\begin{equation}\label{eq:27}
    H(t) = -\frac{1}{2} \nabla^2 - \frac{1}{\left|\mathbf{r}_\mathrm{T}\right|} - \frac{1}{\left|\mathbf{r}_\mathrm{P}(t)\right|}.
\end{equation}
Here,  the electron-nucleus distances are labeled as $\mathbf{r}_\mathrm{T} = \mathbf{r}$ and $\mathbf{r}_\mathrm{P}=\mathbf{r}-\mathbf{R}(t)$.

For the energy regime under consideration, excitation and ionization processes are significantly weaker than the resonant $1s$-$1s$ transfer~\cite{schultz2023data}. This allows us to employ the two-state approximation following established theoretical approaches~\cite{mccarroll1961resonance}, which restricts the atomic orbital space to the minimal basis comprising only the ground states $1s_\text{T}$ and $1s_\text{P}$ centered on the target and projectile nuclei, respectively. This approximation yields $4$ spin orbitals in total and has been demonstrated to reliably reproduce the total charge transfer cross sections for this system~\cite{delos1981theory}. Within this approximation, the second quantization of the Hamiltonian in Eq.~(\ref{eq:27}) can be obtained as:
\begin{equation}
    \begin{aligned}
        H_{\mathrm{sec}}(t) = & \, \sum_{p=0}^3 h_{pp}(t) a_p^\dagger a_p + h_{01}(t) a_0^\dagger a_1 + h_{10}(t) a_1^\dagger a_0 \\
                              &  + h_{23}(t) a_2^\dagger a_3 + h_{32}(t) a_3^\dagger a_2,
    \label{eq:SQH_SM}
    \end{aligned}
\end{equation}
within $4$ spin orbitals by using the same procedure as for the Hamiltonian in Eq.~(\ref{eq:secondQ}). Here, the orbital states of both $\mathrm{H}^{+}$ and $\mathrm{H}$ are obtained by using the minimal STO-3G basis~\cite{hehre1969a, davidson1986basis, zhao2020measurement}. The fermionic operators $a_p^\dagger$ and $a_p$ with $p\in\{0,\,1,\,2,\,3\}$ from Eq.~(\ref{eq:SQH_SM}) are then mapped to qubit operators via the BK transformation, converting the Hamiltonian in Eq.~(\ref{eq:SQH_SM}) to the LCU form of Eq.~(\ref{eq:LCU}) to yield the 4-qubit Hamiltonian:
\begin{equation}
\begin{aligned}
    H_{\mathrm{BK}}(t) = & \, g_0(t)+  g_1(t)X_0  + g_2(t)Y_0 + g_3(t)Z_0 + g_4(t)X_2 \\
                                    + &\,  g_5(t)Y_2 + g_6(t)Z_2 +g_7(t)Z_1X_0 + g_8(t)Z_1Y_0 \\
                                    +& \,   g_9(t)Z_1Z_0+g_{10}(t)Z_3X_2Z_1 + g_{11}(t)Z_3Y_2Z_1\\
                                   +&\, g_{12}(t)Z_3Z_2Z_1,
\end{aligned}
\label{eq:22}
\end{equation}
which can be implemented via universal quantum computers. Here, the relation between the coefficients $g_\gamma(t)$ with $\gamma\in\{0,\cdots,12\}$ and $h_{pq}(t)$ with $p, \,q\in\{0,\,1,\,2,\,3\}$ are provided in Sec.~II of the Supplemental Material. It is worth noting that the resulting qubit Hamiltonian in Eq.~(\ref{eq:22}) exhibits complex time-dependent coefficients that evolve continuously with the internuclear distance, rather than simple linear or step-function time dependence. This non-trivial time dependence provides a rigorous test case for evaluating the capability of quantum algorithms in handling realistic time-dependent dynamics. Complete encoding details are provided in Sec.~II of the Supplemental Material. With the Hamiltonian encoding established in Eq.~(\ref{eq:22}), we proceed to apply variational quantum time evolution algorithms to simulate the collision dynamics.

For the QAS  algorithm, the time-dependent wavefunction ansatz in Eq.~(\ref{eq:QASansatz}) is specified to
\begin{equation}\label{eq:30-1}
    |\Psi(\boldsymbol{\alpha}(t))\rangle=\sum_{i=0}^3\alpha_i(t)\, U_i|1011\rangle,
\end{equation}
which corresponds to the Hamiltonian in Eq.~(\ref{eq:22}).  Here, the $4$ qubit basis states are generated by operators $U_i \in \{I,\, X_0,\, X_2,\, X_2X_0\}$.
Substituting the encoded Hamiltonian of Eq.~(\ref{eq:22}) and the variational ansatze in Eq.~(\ref{eq:30-1})  into McLachlan's variational principle yields ODEs governing parameter dynamics, as presented in Eq.~(\ref{eq:EOMqas}). The evolution of the complex parameters $\boldsymbol{\alpha}(t)$ requires measurements of the overlap matrix $A_{ij} = \langle\varphi_0|U^\dagger_i U_j|\varphi_0\rangle$ and the coupling matrices $D_{\gamma,ij} = \langle\varphi_0|U^\dagger_i H_\gamma U_j|\varphi_0\rangle$. Here, $H_\gamma$ are the Pauli strings extracted from Eq.~(\ref{eq:22}) as defined in Eq.~(\ref{eq:LCU}). Hereafter, all of the algorithms are initialized to the qubit state $|\varphi_0\rangle = |1011\rangle$, which represents the initial electronic configuration with the single electron localized on the hydrogen target.

Similarly, for the AVQDS  algorithm, the ansatz in Eq.~(\ref{eq:AVQDSansatz}) is specified to
\begin{equation}\label{eq:31-1}
    |\Psi(\boldsymbol{\theta}(t))\rangle = \prod_{k=0}^{N_\theta-1} e^{-i\theta_k(t) U_k}|1011\rangle,
\end{equation}
for the proton-hydrogen collision, and can be constructed from a dynamically adapted parameterized circuit shown in Figs.~\ref{fig: circuits}(b) and (c). The operators $U_k$ in Eq.~(\ref{eq:31-1}) are adaptively selected from the set $\mathbb{U}_4$ of the complete $4$-qubit Pauli string defined as in Eq.~(\ref{eq:U}) with $p$ taken from $0$ to $3$. Although the AVQDS ansatz grows dynamically, the total number of selected parameters remains bounded at $N_\theta\le2$ throughout all proton-hydrogen simulations in our study. Substituting the encoded Hamiltonian of Eq.~(\ref{eq:22}) and the variational ansatze in Eq.~(\ref{eq:31-1})  into McLachlan's variational principle yields ODEs governing parameter dynamics, as presented in Eq.~(\ref{eq:EOMavqds}).  For the AVQDS  algorithm, evolution of the real parameters $\boldsymbol{\theta}(t)$ depends on the metric tensor $A^\mathrm{R}_{kl} = \mathrm{Re}\langle\Psi(\boldsymbol{\theta}(t))|U_k^\dagger U_l|\Psi(\boldsymbol{\theta}(t))\rangle$ and the vector $C^\mathrm{I}_{\gamma,k} = \mathrm{Re}\langle\Psi(\boldsymbol{\theta}(t))|U_k^\dagger H_\gamma|\Psi(\boldsymbol{\theta}(t))\rangle$.

The hybrid quantum-classical protocol proceeds as follows: the quantum processor evaluates the required matrix elements through measurement of appropriate observables using the quantum circuits as shown in Fig.~\ref{fig: circuits}, and the classical processor integrates the resulting ODEs to update the variational parameters, thereby obtaining the quantum state $|\Psi(t)\rangle\equiv|\Psi(\boldsymbol{\alpha}(t))\rangle$ for QAS or $|\Psi(t)\rangle\equiv |\Psi(\boldsymbol{\theta}(t))\rangle$ for AVQDS throughout the collision dynamics. Applying this hybrid framework to the proton-hydrogen collision system, our simulations proceed through systematic sampling of the collision parameter space. For each collision energy $E$ in the 1--25~keV range, we simulate 500 uniformly distributed impact parameters up to $b_{\max} = 10\,\text{a.u.}$, with collision trajectories evolved over $vt = 30\,\text{a.u.}$ until the convergence of variational parameters is achieved. From the resulting parameterized state $|\Psi(t)\rangle\equiv|\Psi(\boldsymbol{\alpha}(t))\rangle$ or $|\Psi(t)\rangle\equiv |\Psi(\boldsymbol{\theta}(t))\rangle$ via either QAS or AVQDS  algorithm, we calculate the time-dependent charge transfer probability $P(t)$ and extract the asymptotic charge transfer probability $P(b) \equiv P(t\to\infty)$ from its long-time behavior. These impact parameter-dependent probabilities are then integrated to obtain the charge transfer cross section $\sigma(E)$. Detailed dynamics simulations, accuracy analyses, and cross section results are presented in the following sections.

\subsection{Quantum Simulations for Charge Transfer Processes}

We now present our quantum simulation results for proton-hydrogen collisions using the hybrid quantum-classical computing framework. The calculations were performed on a noiseless quantum simulator to demonstrate the feasibility of digital quantum computation for complex collision problems. Our results reproduce experimental charge transfer cross sections and provide insights into the trade-offs between computational precision and quantum resource requirements.

%%%%%%%%%%%%%%%%%%%%%%%%%%%%%%%%%%%%%%%%%%%%%%%%%%%%%
\begin{figure}
    % \vspace*{11pt}
    \centering
    \includegraphics[width=1\linewidth]{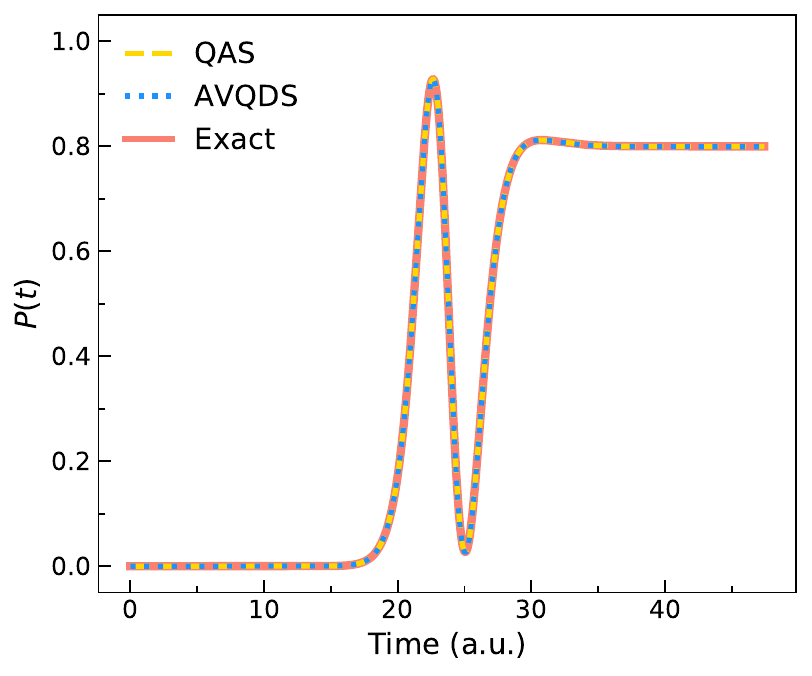}
    \caption{Time-dependent charge transfer probability for $\mathrm{H}^+ + \mathrm{H}(1s)$ collision at $E = 10\,\text{keV}$ and $b = 1.6\,\text{a.u.}$ Results from QAS and AVQDS approaches show excellent agreement with the exact solution obtained by numerically integrating the time-dependent Schr{\"o}dinger equation, with all three curves overlapping.}
    \label{fig: P(t)1}
\end{figure}
%%%%%%%%%%%%%%%%%%%%%%%%%%%%%%%%%%%%%%%%%%%%%%%%%%%%%

\subsubsection{Charge transfer dynamics}

To demonstrate the capability of our hybrid quantum-classical computing framework in modeling atomic collisions, we numerically simulate the $1s$-$1s$ charge transfer dynamics in the proton-hydrogen collision. The time-dependent charge transfer probability for the electron being captured into the traveling $1s$ state of the projectile is denoted by
\begin{equation}
    P(t) = \sum_{\sigma = \uparrow, \downarrow} \left| \langle \Psi_{1s_{\mathrm{P}},\sigma} | \Psi(t) \rangle \right|^2,
\label{eq:Pt}
\end{equation}
where  two possible final states $|\Psi_{1s_\mathrm{P}\uparrow}\rangle$ and $|\Psi_{1s_\mathrm{P}\downarrow}\rangle$ of the  projectile  correspond to $|1010\rangle$ and $|1000\rangle$  in the BK qubit representation, respectively. The time-dependent  state $|\Psi(t)\rangle\equiv|\Psi(\boldsymbol{\alpha}(t))\rangle$ or $|\Psi(t)\rangle\equiv |\Psi(\boldsymbol{\theta}(t))\rangle$ is obtained by either QAS or AVQDS algorithm.

The time evolution of the charge transfer probability $P(t)$ at a collision energy of $E = 10\,\text{keV}$ and an impact parameter of $b = 1.6$~a.u. is simulated and shown in Fig.~\ref{fig: P(t)1}. For comparison, we obtain the exact solution by numerically integrating the time-dependent Schr{\"o}dinger equation governed by the Hamiltonian in Eq.~(\ref{eq:22}) using the QuTiP package~\cite{johansson2012qutip}. The results from both variational quantum time evolution algorithms, QAS and AVQDS, demonstrate perfect agreement with this exact solution, validating the accuracy of our quantum algorithmic approach. Initially, as the target and the projectile are separated by 15~a.u. (far exceeding the Bohr radius), the charge transfer probability $P(t)$ remains zero over a period of time. As the projectile approaches, the increasing Coulomb interaction induces oscillations in the charge transfer probability, which is a characteristic feature of resonant collision processes~\cite{ stich1983time}. Beyond the closest approach, the charge transfer probability stabilizes at approximately $80\%$ as the Coulomb interaction weakens with increasing internuclear separation, corresponding to the asymptotic charge transfer probability $P(b)$ at this impact parameter.

In the numerical implementation of ODEs using classical computers, we employ different integration strategies for the QAS and AVQDS algorithms, based on their distinct parameter scaling behaviors. In the QAS framework, parameter reduction is completed during the preprocessing, resulting in a fixed parameter dimensionality of four throughout the time evolution. This fixed dimensionality enables us to employ the Adams predictor-corrector integrator~\cite{brown1989vode} with adaptive time steps, where integration accuracy is controlled through absolute and relative error tolerances. In contrast, the AVQDS approach adaptively adjusts the number of variational parameters during the simulation based on the McLachlan distance, leading to time-dependent dimensionality of the ODEs.

When the adaptive-step integrators used in QAS are applied to ODEs with time-varying dimensionality in AVQDS, we find that significant numerical instabilities emerge at these parameter dimension transition points. For such discontinuous dimensional changes, the Euler integrator demonstrates better numerical stability despite its lower theoretical accuracy, owing to its fixed time step and linear prediction characteristics. This integration method incurs a local truncation error of $\mathcal{O}(\Delta t^2)$ at each step. For large impact parameters or high collision energies, these errors accumulate and may lead to a slight drift in $P(b)$. Smaller $\Delta t$ values generally yield results closer to those from the adaptive-step integration, but also incur higher quantum measurement overhead. In our work, we choose a small time step of $\Delta t = 0.005\,\text{a.u.}$ and a strict McLachlan threshold $L_{\text{cut}} = 10^{-8}$ to ensure high-accuracy results for theoretical validation. Under these settings, the AVQDS ansatz achieved a maximum of two variational parameters in the present simulation.

\subsubsection{Quantum state fidelity analysis}

\begin{figure}
    \centering
    \includegraphics[width=1\linewidth]{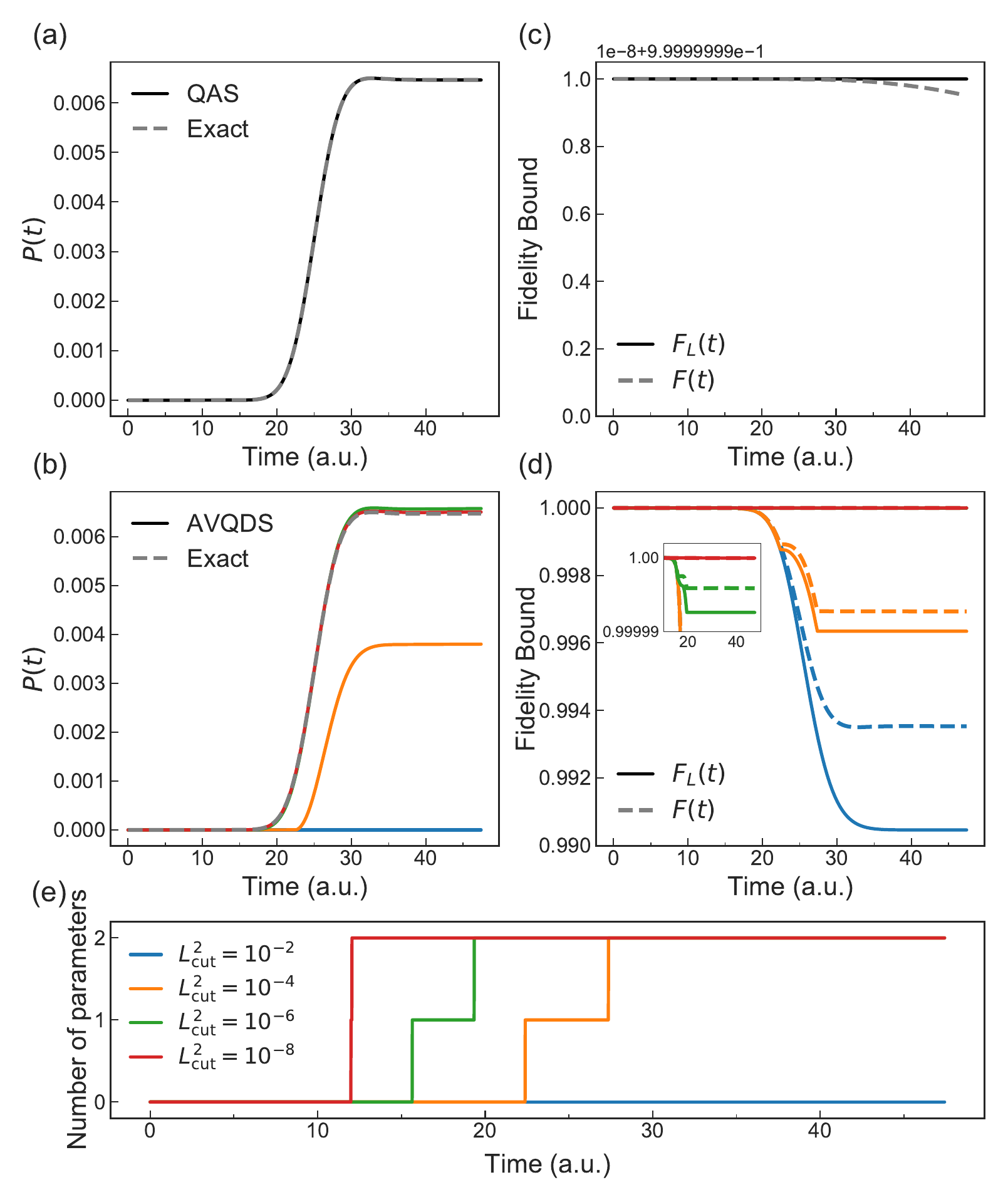}
    \caption{Time-dependent charge transfer processes at $E = 10\,\text{keV}$ and $b = 6.0\,\text{a.u.}$ (a) Charge transfer probability from the QAS approach (solid line) and numerically exact results (dashed line). (b) Charge transfer probabilities from the AVQDS approach using different McLachlan distance thresholds $L^2_\text{cut}$ (solid lines) compared to the same numerically exact results (dashed line) in (a), with $L^2_\text{cut} = 10^{-2}$ (blue), $10^{-4}$ (orange), $10^{-6}$ (green), and $10^{-8}$ (red). (c) and (d) Variational fidelity bounds (solid lines) and quantum state fidelities (dashed lines) for QAS and AVQDS with different thresholds. (e) Number of parameters required in AVQDS as a function of time for different thresholds.}
    \label{fig: P(t)5}
\end{figure}

To evaluate the accuracy of the simulation throughout the evolution, we analyze the quantum state fidelity, defined as
\begin{equation}
    F(t) = \left| \langle \Psi_{\text{exact}}(t) | \Psi(t) \rangle \right|^2.
\label{eq:fidelity}
\end{equation}
Here, $|\Psi_\text{exact}(t)\rangle$ represents the numerically exact solution governed by the Hamiltonian in Eq.~(\ref{eq:22}), and is obtained via the QuTiP package~\cite{johansson2012qutip}.  However,  the parameterized state $|\Psi(t)\rangle\equiv|\Psi(\boldsymbol{\alpha}(t))\rangle$ or $|\Psi(t)\rangle\equiv |\Psi(\boldsymbol{\theta}(t))\rangle$ is obtained by either QAS or AVQDS algorithm. A comparison of the QAS and AVQDS results at $E = 10\,\text{keV}$ and $b = 6.0\,\text{a.u.}$ is presented in Fig.~\ref{fig: P(t)5}, focusing on the quantum state fidelity $F(t)$ and the influence of the McLachlan distance threshold $L^2_\text{cut}$.

%%%%%%%%%%%%%%%%%%%%%%%%%%%%%%%%%%%%%%%%%%%%%%%%%%%%%
%%%%%%%%%%%%%%%%%%%%%%%%%%%%%%%%%%%%%%%%%%%%%%%%%%%%%

As shown in Figs.~\ref{fig: P(t)5}(a) and (b), both QAS and AVQDS produce charge transfer probabilities with much smoother evolution compared to the oscillatory behavior in Fig.~\ref{fig: P(t)1}. This reduced oscillatory behavior reflects the weak Coulomb interactions at large impact parameters, which suppress both oscillations and the overall magnitude of charge transfer probabilities~\cite{ludde1981direct}.

The quantum state fidelities, represented by dashed lines in Figs.~\ref{fig: P(t)5}(c) and (d), provide a quantitative evaluation of the accuracy of these simulations. It is evident that the QAS approach not only reproduces the charge transfer probabilities with great agreement, but also maintains a consistently high fidelity throughout the evolution, achieving a final infidelity $\mathcal{I}=1-F(t\to\infty)$ of $\mathcal{O}(10^{-10})$. In contrast, while the AVQDS exhibits similarly smooth charge transfer probabilities, both the charge transfer probability and quantum state fidelity strongly depend on the McLachlan distance threshold $L^2_\text{cut}$. This threshold serves as a key control parameter in the AVQDS approach, governing the adaptive evolution of the variational ansatz. When the McLachlan distance $L^2$ exceeds $L^2_\text{cut}$ during the evolution, new operators are incorporated into the quantum circuit to enhance the ansatz's expressibility capability. This adaptive mechanism is further illustrated in Fig.~\ref{fig: P(t)5}(e), which displays the time evolution of the number of parameters in the variational ansatz during AVQDS simulations.

The choice of the McLachlan distance threshold $L^2_\text{cut}$ requires balancing simulation accuracy and computational resources. As illustrated in Figs.~\ref{fig: P(t)5}(b), (d), and (e), a smaller threshold enables earlier incorporation of necessary operators, resulting in improved simulation fidelity. However, we also observe that thresholds below $10^{-10}$ can lead to simulation failure as numerical precision limitations prevent achieving such stringent convergence criteria. On the other hand, larger thresholds greater than $10^{-2}$ may result in insufficient ansatz expressibility, particularly in regimes with small charge transfer probabilities. Through a systematic investigation of various collision parameters, the optimal range of $L^2_\text{cut}$ for the proton-hydrogen collision problem is found to be $10^{-8}$ to $10^{-4}$. To ensure high-fidelity results across all parameter regimes in subsequent cross section calculations, we adopt $L^2_\text{cut}=10^{-8}$ hereafter. This threshold yields an optimal final infidelity of $\mathcal{O}(10^{-7})$ for AVQDS results.

%%%%%%%%%%%%%%%%%%%%%%%%%%%%%%%%%%%%%%%%%%%%%%%%%%%%%
\begin{figure*}
    % \centeringh
    \includegraphics[width=1\textwidth]{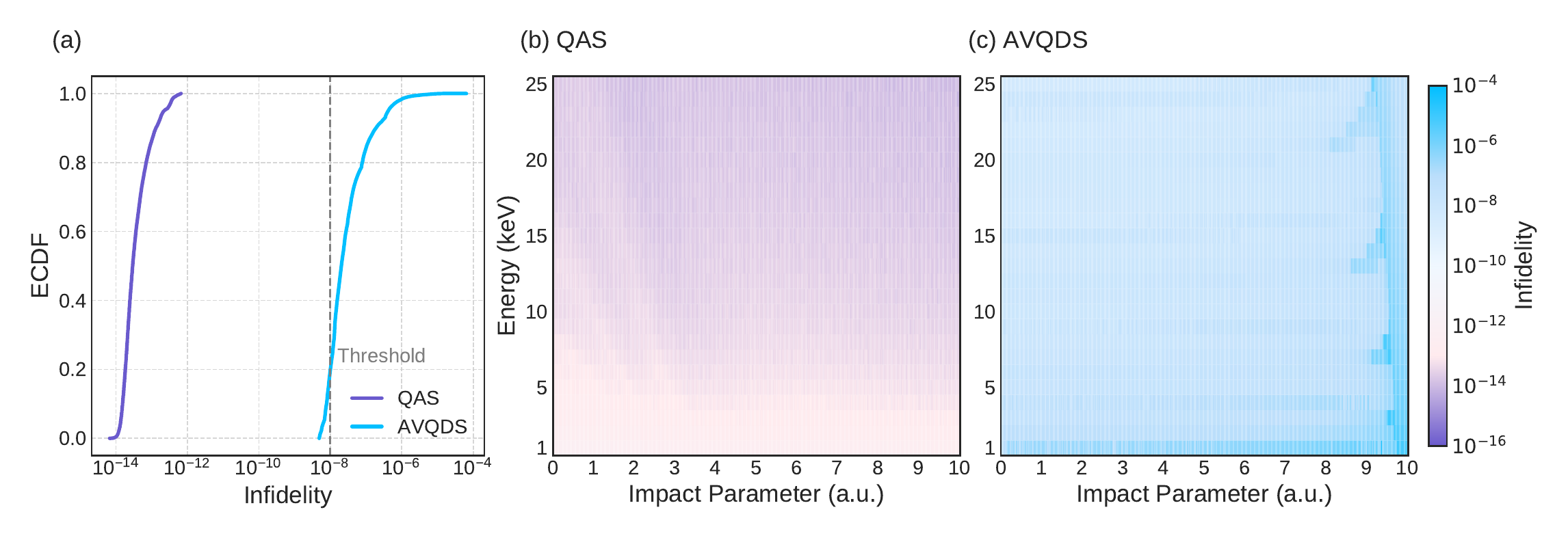}
    \caption{(a) Empirical cumulative distribution function (ECDF) of infidelity across all $(E, b)$ simulation points. Infidelity values below $10^{-8}$ may not account for numerical errors outside $F_L$ as discussed earlier, but this does not affect the conclusion that both methods achieve sufficient precision. (b) and (c) Infidelity heatmaps for QAS and AVQDS, evaluated using the variational fidelity bound $F_L$ as a function of impact parameter and collision energy. These results demonstrate the utility of $F_L$ as a practical diagnostic for assessing simulation accuracy throughout the entire parameter space.}
    \label{fig: heat}
\end{figure*}
%%%%%%%%%%%%%%%%%%%%%%%%%%%%%%%%%%%%%%%%%%%%%%%%%%%%%

Notably, although both QAS and AVQDS approaches attempt to minimize the number of parameters required for accurate time evolution, they differ fundamentally in their resource requirements. The QAS approach maintains constant circuit depth of $\mathcal{O}(1)$, which is independent of parameter numbers $N_\alpha$ and requires only initial measurements. In contrast, the AVQDS approach requires measurements at each time step to update the ansatz and evaluate the McLachlan distance, leading to increased quantum computational demands. Specifically, the circuit depth scales as $\mathcal{O}(N_\theta)$ with the number of variational parameters, while the total number of measurements scales polynomially with $N_\theta$ and the number of time steps $N_t$. A lower threshold, while potentially improving accuracy, leads to earlier parameter incorporation and thus increases the average circuit depth for each simulation. Similarly, as previously discussed, decreasing the evolution step size leads to a significant increase in the total measurement requirements. These distinctions become particularly significant when considering experimental implementation, where circuit depth optimization, measurement overhead, and experimental noise may favor a more relaxed threshold $L^2_\text{cut}$ and optimized time step $\Delta t$, despite the theoretical capability for higher precision.

\subsubsection{Parameter-space variational fidelity bound}

While fidelity is widely used as a benchmark for assessing simulation accuracy, the variational fidelity bound based on the McLachlan distance $L^2(t)$ provides an alternative accuracy criterion that does not require additional computational resources for exact state calculations~\cite{lubich2005variational, martinazzo2020local, zoufal2023error, gacon2024variational}. Instead, this bound only leverages results from the variational calculations themselves, with detailed derivation provided in Sec.~I of the Supplemental Material. The variational fidelity bound is defined by
\begin{equation}
    F_{L}(t) \equiv \left(1 - \frac{\epsilon(t)^2}{2}\right)^2 \leq F(t) ,
\label{eq:fidelityLowerBound}
\end{equation}
where $\epsilon(t)= \int_0^t \|L(\eta)\|\: \mathrm{d}\eta$ is the cumulative McLachlan error, with $L(\eta)\equiv\sqrt{L^2(\eta)}$ obtained from the McLachlan distance $L^2(\eta)$ measured during the variational evolution. $F(t)$ is defined in Eq.~(\ref{eq:fidelity}).  We additionally plot the variational fidelity bounds for different thresholds as solid lines in Figs.~\ref{fig: P(t)5}(c) and (d). These numerical results confirm that the variational fidelity bound consistently serves as a lower bound to the quantum state fidelity in most cases.

%%%%%%%%%%%%%%%%%%%%%%%%%%%%%%%%%%%%%%%%%%%%%%%%%%%%%
\begin{figure*}
    % \vspace*{8pt}
    \centering
    \includegraphics[width=0.9\textwidth]{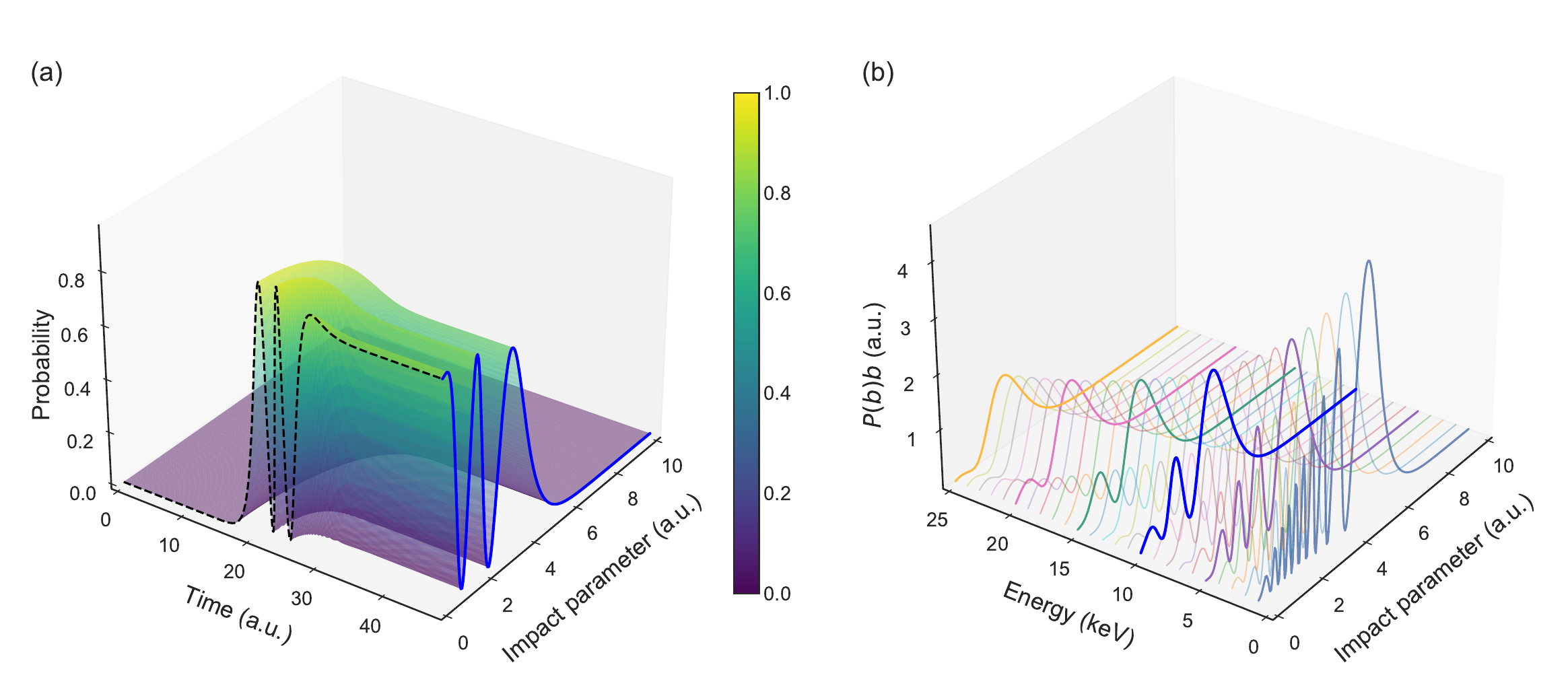}
    \caption{Charge transfer dynamics in H$^+$ + H(1s) collisions calculated using the QAS approach.
    (a) Charge transfer probabilities versus time and impact parameters at $E=10\,\text{keV}$. Two representative curves are shown: the time-dependent charge transfer probability $P(t)$ (black dashed line) at a small impact parameter $b=0.02\,\text{a.u.}$ and the asymptotic charge transfer probability $P(b)$ (solid royal blue line) as a function of impact parameter used in cross section calculations. (b) Integrand $P(b)b$ as a function of impact parameter at collision energies from 1~keV to 25~keV. The royal blue curve represents results for the same $E=10\,\text{keV}$ collision energy.}
    \label{fig: 3D}
\end{figure*}
%%%%%%%%%%%%%%%%%%%%%%%%%%%%%%%%%%%%%%%%%%%%%%%%%%%%%

Despite the theoretical expectation that $F_L(t) \leq F(t)$, minor deviations may occur in extremely high-precision regimes when the McLachlan distance becomes vanishingly small (e.g., below $10^{-8}$) and approaches the magnitude of numerical integration errors. Similar effects are also observed in QAS as shown in Fig.~\ref{fig: P(t)5}(c). Although QAS does not explicitly use $L_\mathrm{cut}^2$ for accuracy control, it is still based on the McLachlan formalism and achieves extremely low $L^2$ values due to its high-order ODE solvers. This phenomenon reflects a fundamental shift in error dominance: from variational errors to numerical integration errors. This occurs because $F_L(t)$ is derived locally from the variational ansatz and remains insensitive to errors outside the variational subspace, while $F(t)$, as a global observable, captures accumulated truncation and rounding errors over the full evolution. As a result, $F(t)$ may drift slightly downward while $F_L(t)$ remains unaffected, producing the observed minor inequality violations. Nevertheless, these minor deviations occur only in ultra-high-precision regimes and remain numerically insignificant for practical collision simulation applications.

Building on this, the variational fidelity bound $F_L(t)$ remains a reliable and computationally efficient alternative to quantum state fidelity evaluations. Our analysis shows that computing the exact state evolution usually incurs a time cost at least an order of magnitude higher than variational quantum approaches. This computational overhead makes direct quantum state fidelity analysis impractical for parameter-space sweeps that are commonly required in complex dynamics, such as cross section calculations in collision problems. Therefore, by continuously monitoring $F_{L}(t)$, we can efficiently ensure high fidelity throughout the entire simulation, providing a scalable diagnostic tool for variational quantum dynamics without requiring access to the exact state.

To assess the overall performance of our collision simulations, we construct infidelity heatmaps spanning the full collision parameter space $(E,b)$, based on the variational fidelity bound $F_L(t)$. The collision parameter space covers collision energies from 1~keV to 25~keV and impact parameters from 0.02~a.u. to 10~a.u. To provide a global perspective on simulation accuracy, we first examine the statistical distribution of infidelities derived from the heatmap data in Fig.~\ref{fig: heat}(a). The empirical cumulative distribution function shows that both QAS and AVQDS achieve high fidelity over the entire parameter space, with QAS predominantly concentrated in the $10^{-14}$--$10^{-12}$ range and AVQDS in the $10^{-8}$--$10^{-6}$ range.

The parameter-space distribution of these infidelities is further detailed in the heatmaps of Figs.~\ref{fig: heat}(b) and (c), which serve as comprehensive visualization tools for rapid identification of accuracy ranges, stability boundaries, and convergence issues. The detailed patterns across the parameter space show that these infidelities remain robust even in the presence of numerical deviations discussed earlier, which do not compromise the observed accuracy trends. We note that a small fraction of AVQDS simulations, most notably at $E = 1\,\text{keV}$, failed to converge at certain impact parameters due to the McLachlan threshold not being met. These points were patched using nearest-neighbor interpolation. These threshold-induced convergence failures can be mitigated by adjusting the engineering parameters such as the evolution time step or the McLachlan threshold, and were not encountered in QAS, which does not depend on threshold-based accuracy control. Overall, both methods achieve high precision within their respective computational frameworks, demonstrating the consistent and reliable performance of variational quantum time evolution algorithms in achieving high-fidelity dynamics across a broad parameter space and supporting their applicability for efficient and accurate quantum dynamics simulations.

\subsubsection{Charge transfer cross sections}

We now compute the charge transfer cross sections for proton-hydrogen collisions over the energy range of 1--25~keV employing our hybrid quantum-classical computing framework. The cross sections are determined as follows:
\begin{equation}
    \sigma_{}(E) = 2\pi \int_0^{\infty} P(b) b \, db.
\label{eq:totalCrossSection}
\end{equation}%\mathrm{total}
Here, $P(b)$ denotes the asymptotic charge transfer probability extracted from the converged evolution at each impact parameter, as depicted in Fig.~\ref{fig: 3D}(a). At the representative energy of 10~keV in Fig.~\ref{fig: 3D}(a), the asymptotic charge transfer probability decreases exponentially for impact parameters beyond 5.0~a.u., consistent with the suppressed probability magnitude observed in Figs.~\ref{fig: P(t)5}(a) and (b). We further compute $P(b)b$ as a function of impact parameter for collision energies ranging from 1~keV to 25~keV, with results shown in Fig.~\ref{fig: 3D}(b). At lower energies, these curves exhibit more pronounced oscillations with larger amplitudes, particularly in the intermediate impact parameter region. This behavior indicates that the intermediate impact parameter region contributes dominantly to the cross section. These results not only quantify the contributions from each impact parameter but also reveal characteristic features of charge transfer dynamics in resonant collisions~\cite{cheshire1968proton, ludde1982electron}.

\begin{figure}
    \centering
    \includegraphics[width=1\linewidth]{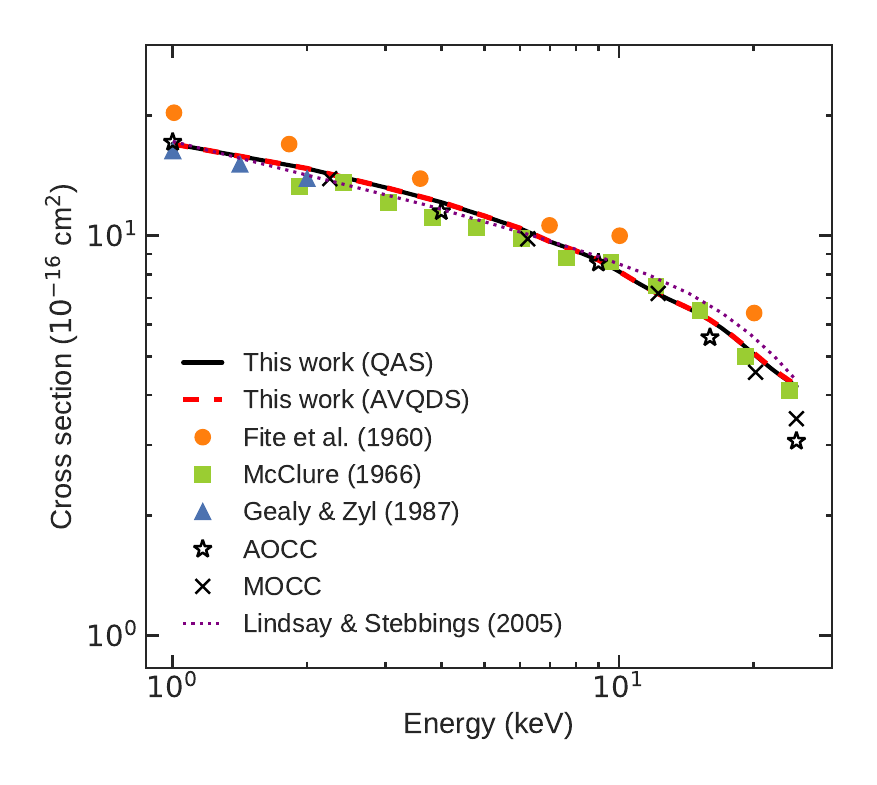}
    \caption{The charge transfer cross sections for $\mathrm{H}^+ + \mathrm{H}(1s)$ collisions. Our results from the QAS approach are shown with a solid black line and the AVQDS approach are shown with a dashed red line. They are compared with the experimental data from Fite \emph{et al.}~\cite{fite1960ionization}, McClure~\cite{mcclure1966electron} and Gealy and van Zyl~\cite{gealy1987cross}, as well as the theoretical results from AOCC~\cite{agueny2019electron} and MOCC~\cite{harel1998cross}, along with the recommended value from Lindsay and Stebbings~\cite{lindsay2005charge}.}
    \label{fig: cross sections}
\end{figure}

In Fig.~\ref{fig: cross sections}, cross sections calculated by QAS and AVQDS approaches are presented and compared with available experimental data and other theoretical results obtained via classical computers. The experimental data include measurements by Fite \emph{et al.}~\cite{fite1960ionization}, McClure~\cite{mcclure1966electron}, and Gealy and van Zyl~\cite{gealy1987cross}. Previous theoretical results based on AOCC~\cite{agueny2019electron} and MOCC~\cite{harel1998cross} calculations and empirical fit results from Lindsay and Stebbings~\cite{lindsay2005charge} are also included. The QAS and AVQDS calculations exhibit very good agreement with both experimental and theoretical results across the entire energy range we considered. Both variational calculations clearly capture the characteristic increase in total charge transfer cross sections with decreasing collision velocity,  which is a signature feature of resonant collisions~\cite{bransden1992charge}.

%%%%%%%%%%%%%%%%%%%%%%%%%%%%%%%%%%%%%%%%%%%%%%%%%%%%%
%%%%%%%%%%%%%%%%%%%%%%%%%%%%%%%%%%%%%%%%%%%%%%%%%%%%%

With comparison to experimental data, our calculations agree well with the existing experimental measurements, particularly with the data of McClure~\cite{mcclure1966electron} and Gealy and van Zyl~\cite{gealy1987cross}. The calculated cross sections successfully reproduce both the magnitude and the energy dependence of the experimental data. To quantitatively assess this agreement, we present the relative error of our computed cross sections from the experimental results of McClure~\cite{mcclure1966electron} and Gealy and van Zyl~\cite{gealy1987cross} in Table~\ref{tab:CScomparison}. For clarity, Table~\ref{tab:CScomparison} only shows the QAS results for its higher precision. Detailed cross section data for both  QAS and AVQDS are available in Sec.~IV of the Supplemental Material. The average relative error across the studied energy range is $5.93\%$, with a maximum deviation of 11.7$\%$ at 1.92~keV and a minimum deviation of 2.13$\%$ at 15.2~keV. Notably, for most energy points, our theoretical predictions fall within or close to the reported experimental uncertainties of $\pm5\%$ in McClure~\cite{mcclure1966electron} and $\pm17\%$ in Gealy and van Zyl~\cite{gealy1987cross}. Furthermore, the slight overestimation observed in our theoretical results compared to experimental measurements may be attributed to the two-state approximation used in our calculations. This approximation implicitly redirects the probabilities of excitation and ionization processes into the charge transfer probability, potentially leading to enhanced cross section values.

\begin{table}
\caption{Comparison of QAS cross sections and experimental data from McClure~\cite{mcclure1966electron} and Gealy and van Zyl~\cite{gealy1987cross} in units $10^{-16}$ cm$^2$. The relative error is calculated as $\varepsilon_{\text{QAS}}=|\sigma_{\text{QAS}} - \sigma_{\text{exp}}| / \sigma_{\text{exp}} \times 100\%$. Here, the subscripts ``$\text{exp}$" and ``$\text{QAS}$" denote the results of experiments and  QAS theoretical calculations, respectively.}
\label{tab:CScomparison}
\begin{ruledtabular}
\begin{tabular}{ccccc}
\textrm{$E$ (keV)} & \textrm{$\sigma_{\text{exp}_1}$\footnote{Experimental cross sections $\sigma_{\text{exp}_1}$ from Ref.~\cite{gealy1987cross}.}} & \textrm{$\sigma_{\text{exp}_2}$\footnote{Experimental cross sections $\sigma_{\text{exp}_2}$ from Ref.~\cite{mcclure1966electron}.}} & \textrm{$\sigma_{\text{QAS}}$} & $\varepsilon_\text{QAS}$\\
\hline
1.00  &16.3&       & 17.0  & 4.17\%  \\
1.41  &15.1&       & 15.8  & 4.87\%  \\
1.92  &    & 13.3  & 14.9  & 11.7\%  \\
2.00  &13.9&       & 14.7  & 6.00\%  \\
2.41  &    & 13.6  & 14.0  & 3.05\%  \\
3.04  &    & 12.1  & 13.1  & 8.65\%  \\
3.82  &    & 11.1  & 12.3  & 11.2\% \\
4.80  &    & 10.5  & 11.3  & 7.95\%  \\
6.05  &    & 9.85  & 10.4  & 5.54\%  \\
7.62  &    & 8.8   & 9.35  & 6.20\%  \\
9.60  &    & 8.6   & 8.37  & 2.73\%  \\
12.1  &    & 7.5   & 7.20  & 4.04\%  \\
15.2  &    & 6.5   & 6.36  & 2.13\%  \\
19.2  &    & 5.0   & 5.27  & 5.49\%  \\
24.1  &    & 4.1   & 4.32  & 5.30\%  \\
\end{tabular}
\end{ruledtabular}
\end{table}

When compared with previous theoretical studies, our results demonstrate consistent agreement across most of the energy spectrum, with slight deviations only at energies above 16~keV. While well-established close-coupling methods, such as AOCC and MOCC, employ substantially larger basis sets, our quantum simulation results show better agreement with experimental data and empirical fitting curves in this energy range. This demonstrates that our variational quantum dynamics simulations successfully model the charge transfer mechanisms with higher accuracy using only computationally efficient minimal basis sets. The accurate performance of our quantum approaches validates their capability in simulating electron transfer dynamics and cross sections in proton-hydrogen atom collisions for both regimes of low and intermediate energies. Overall, our calculations with the hybrid quantum-classical computing framework successfully reproduce the experimental charge transfer cross sections across the studied energy range.

%%%%%%%%%%%%%%%%%%%%%%%%%%%%%%%%%%%%%%%%%%%%%%%%%%%%

\section{Discussions and Conclusions}\label{sec:V}

In this work, we have developed a hybrid quantum-classical computing framework for simulating ion-atom collision dynamics using two variational quantum time evolution algorithms. Within this framework, we implemented and benchmarked two distinct variational algorithms by using a typical example on proton-hydrogen collisions across the $1$~keV to $25$~keV energy regime. Both algorithms accurately capture the time-dependent charge transfer dynamics, maintaining high fidelity throughout the entire evolution. The resulting charge transfer cross sections, calculated using only the minimal basis, show very good agreement with experimental measurements. Based on the QAS results, the average relative error across the studied energy range is 5.93\%. The average relative error for the AVQDS results is 5.99\%. Additionally, we introduce the variational fidelity bound heatmap as a practical and scalable diagnostic tool for assessing simulation accuracy across different parameters without requiring access to the exact quantum state.

Beyond the specific case of proton-hydrogen collisions, the proposed variational framework is applicable to a broad class of time-dependent many-body quantum dynamics problems that involve multiple electrons and heavier ions, providing a systematic and hardware-compatible approach for simulating such processes on near-term quantum devices. The variational fidelity bound heatmap introduced in this work offers a standardized diagnostic tool for evaluating the accuracy of variational quantum simulations in diverse dynamical applications. Furthermore, this study demonstrates the potential of applying variational quantum algorithms to quantum scattering and charge transfer problems, which remain relatively unexplored within the context of quantum computing. These findings open new directions for extending variational approaches to more complex and realistic time-dependent problems.

In extending this framework to broader applications, it is essential to understand the complementary strengths of these variational algorithms and how they may be optimized based on specific accuracy and resource constraints. The QAS approach offers exceptional numerical precision with minimal quantum hardware requirements, making it particularly suitable as a reference approach. In contrast, the AVQDS approach incorporates more quantum computational components via parametrized quantum circuits, suggesting greater potential for quantum acceleration as hardware matures. Although it shows slightly lower numerical precision than QAS, AVQDS provides a more direct pathway toward quantum advantage. This complementarity highlights the importance of selecting variational strategies based on available quantum resources and precision demands.

Several optimization pathways remain to improve the algorithm implementations. For the QAS approach, one promising direction involves reducing the number of qubits and variational parameters by leveraging the absence of spin and angular momentum coupling in the underlying Hamiltonian. This simplification may enable more compact matrix representations and reduce the overall quantum resource requirements. For the AVQDS approach, which inherently avoids parameter redundancy, measurement overhead can be significantly reduced by incorporating adaptive integrators with warm-starting strategies at parameter dimension transition points in the ODE system, while still maintaining integration accuracy. We mention that this work only considers an noiseless ideal quantum simulator, the environmental effect on the simulation results of the quantum simulator will be further studied in the future works.

In conclusion, this study establishes ion-atom collisions as a paradigmatic class of problems for NISQ-era quantum computing, while advancing the understanding of non-equilibrium quantum dynamics. By successfully mapping collision problems onto the hybrid quantum-classical computing frameworks, we enable systematic exploration of quantum advantage in atomic and molecular physics. Our work also validates the effectiveness of variational quantum approaches for time-dependent physical problems, with possible extensions to resource-efficient ansatz design, higher-energy regimes, and more complex many-electron systems, ultimately enabling applications in nuclear, plasma, and astrophysical systems using near-term quantum devices.

\begin{acknowledgments}
This work was supported by Innovation Program for Quantum Science and Technology with Grant No. 2021ZD0300201 and the National Natural Science Foundation of China with Grants No.~12374483 and No.~92365209.
\end{acknowledgments}

%%%%%%%%%%%%%%%%%%%%%%%%%%%%%%%%%%%%%%%%%%%%%%%%%%%%

%%%%%%%%%%%%%%%%%%%%%%%%%%%%%%%%%%%%%%%%%%%%%%%%%%%%

\end{document}